\documentclass[twocolumn]{aastex7}
\usepackage{amsmath,amstext}
\usepackage[T1]{fontenc}
\usepackage{subfigure}
\usepackage{natbib}
\usepackage{sidecap}
\usepackage{enumitem}
\usepackage{appendix}
\usepackage[export]{adjustbox}
\usepackage{caption}
\usepackage{subcaption}
\usepackage{float}
\usepackage{accents}
\usepackage{xfrac}

\newcommand{\DV}{}
\newcommand{\jks}{}

\begin{document}

\title[The km-size barrier around WDs]{Size limits on tidal debris around white dwarfs: the km-size barrier}

\author[0000-0002-1717-2226]{Jordan K. Steckloff}
\affiliation{Department of Aerospace Engineering and Engineering Mechanics, University of Texas at Austin, Austin, TX, USA}
\affiliation{Planetary Science Institute, 1700 E Fort Lowell Rd STE 106, Tucson, AZ 85719, USA}
\email{jsteckloff@psi.edu}

\author[0000-0001-8014-6162]{Dimitri Veras}
\affiliation{Centre for Exoplanets and Habitability, University of Warwick, Coventry CV4 7AL, UK}
\affiliation{Centre for Space Domain Awareness, University of Warwick, Coventry CV4 7AL, UK}
\affiliation{Department of Physics, University of Warwick, Coventry CV4 7AL, UK}
\email{dimitri.veras@aya.yale.edu},

\author[0000-0001-8736-236X]{Kathryn Volk}
\affiliation{Planetary Science Institute, 1700 E Fort Lowell Rd STE 106, Tucson, AZ 85719, USA}
\email{kvolk@psi.edu}

\begin{abstract}

Compact disks of planetary debris orbiting white dwarfs provide a crucial window into our understanding of evolved planetary systems. 
The formation of these disks has been widely modeled with tidal fragmentation of minor planets that are rubble piles with no internal strength. 
However, rubble piles do have non-zero cohesive strength from Van der Waals forces, and here we demonstrate the consequences: breakup of these rubble piles sets a maximum fragment size, and we calculate this size \jks{for water ice, iron, and material densities corresponding to the lunar highlands, Vesta and the Earth}. 
We find that for typical minimum rubble pile strengths of $\sim$10-1000 Pa, the maximum fragment size is as large as small asteroids (0.1-1 km). 
This limit -- the km-size barrier -- also represents the characteristic sizes of tidal fragments. 
Most of the debris mass is contained in fragments of this size. 
Consequently, subsequent disk evolution should first feature a prominent dust-forming process, such as collisional grinding, before Poynting-Robertson drag can significantly shape the final disk. \jks{Further, we find that non-zero internal strength more narrowly radially confines the fragments than in the strengthless case.}  This correction to previous assumptions adds to the growing evidence of the importance of collisions in the formation and evolution of white dwarf debris disks, while also helping to bound the size distribution in these disks for modeling efforts. 

\end{abstract}


\section{Introduction}\label{s:intro}

While the complexity and cadence of new observations of white dwarf planetary systems mount, predictive and explanatory models for these systems need to keep up. 
As this now decades-old field of study matures, basic canonical models of the systems’ dynamical origins have become decreasingly applicable to the increased diversity of detections.
Almost 2,000 \jks{likely} white dwarf planetary systems are now known through the detection of accreted metals in their stellar atmospheres \citep{Williams:2024}. 
The metals cannot be primordial to the star nor arise from passing interstellar clouds; they must arise from molecularly shorn planetary debris \citep[e.g.][]{Zuckerman:2003}.  
These accreted metals are inextricably linked to debris disks which surround the white dwarfs at the approximate location of their Roche radius, emphasizing the importance of understanding the evolution of these disks.

These disks are assumed to be common features of white dwarf planetary systems. 
On the order of 100 of these disks can currently be detected \citep[e.g.][]{Zuckerman:1987,Gansicke:2006,Farihi:2009,Manser:2020,Xu:2020,Rogers:2024,Farihi:2025,Reach:2025,Wang:2026,Wang:2026b}.
However, nearly every one of the $\sim$2,000 known white dwarfs that are polluted or enriched with metals is assumed to accrete from a disk due to the improbability of a direct collision between a planetary body and a white dwarf photosphere \citep{Brown:2017,McDonald:2021}.

In recent years, the activity levels and locations of these disks have introduced outstanding challenges for models of their formation. 
Almost all white dwarf disks are now known to showcase flux changes on the orders of 1-10\% over timescales ranging from minutes to months \citep{Swan:2020,Noor:2025}, and recent coordinated observations of both the dust and gas in these disks reveal simultaneous activity spikes for both \citep{Rogers:2025}. 
An outburst in \jks{the WD 0145+234 system} \citep{Wang:2019} generated a 1.0 magnitude activity spike between 2018 and 2019, prompting a more detailed look at collisional evolution in these disks \citep{Swan:2021}. 
These activity levels remain largely unexplained, and few investigations have modeled the gas-dust interplay within these systems \citep{Metzger:2012,Bromley:2017,Okuya:2023}.

The locations of these disks, as partially revealed by periodic signatures in photometric transits, have been similarly intriguing. 
Although the first such detections \citep{Vanderburg:2015} conformed to expectations, with periodicities of $\approx$4.5 hours that correspond to the approximate white dwarf Roche radius, subsequent detections have revealed periodicities of 5 \citep{Guidry:2025}, 10 \citep{Vanderbosch:2021}, 15 \citep{Hermes:2025}, and 25 hours \citep{Farihi:2022} as well as $\sim$100 days \citep{Vanderbosch:2020}. 
The variety of theories for how planetary debris has settled at these locations exterior to the Roche radius are mounting \citep{Veras-Kurosawa:2020,Veras:2020, Veras:2022,Kislyakova:2023,Shestakova:2023,Li:2025,Li:2025b,Veras:2025,Veras:2026}, but regardless often feature destruction processes of some kind.

The formation of white dwarf debris disks has traditionally been numerically modeled as  arising from the break-up of a strengthless rubble-pile asteroid \citep{Debes:2012,Veras:2014,Veras:2017,Duvvuri:2020,Malamud:2020,Malamud:2020b, Brouwers:2023}.  
This breakup helps determine the resulting size frequency distribution of the debris, their collisional evolution, their evolution due to radiative drag, the potential production of gas, and ultimately the lifetime of the disk; 
estimates of these lifetimes vary by orders of magnitude \citep{Girven:2012, Veras-Heng:2020,Cunningham:2021,Cunningham:2025}.

However, our knowledge from rubble piles in the solar system demonstrates that they {\it do} have nonzero strength. 
Rubble pile asteroids have cohesive strength due to Van der Waals forces between grains \citep{Sanchez:2014,Scheeres:2018}. 
Estimates for the minimum tensile strengths of planetesimals (asteroids and comets) range from $\sim$1-1000 Pa for comets \citep{Sekanina:1985,Asphaug:1996, Steckloff:2015,Hirabayashi:2016,Attree:2018} and rubble pile asteroids \citep{Scheeres:2018,Arakawa:2020,Brisset:2022,Perry:2022} and from $\sim$kPa - MPa for more consolidated regoliths and monolithic asteroidal materials \citep{Mitchell:1972,Pohl:2020,Chang:2022}. 
Thus,  for fragmentation to occur, tidal forces must not only overcome gravitational forces but also material strength. Despite not often being explicitly modeled in N-body simulations, this interplay between tides, gravity, and strength has been widely acknowledged in the white dwarf planetary system community \citep{Bear:2015,Brown:2017,Manser:2019,O'Connor:2020,Veras:2020,Li:2021,McDonald:2021,Zhang:2021, Brouwers:2022}.

The goal of this Letter is to compute characteristic maximum fragment sizes due to realistic minimum strength values of extrasolar analogs of solar system asteroids, while emphasizing that the longstanding strengthless assumption for any type of minor planet is unrealistic. 
In Section~\ref{s:methods}, we present our physical setup from which the characteristic sizes are computed; this formula is compared to expressions for the widely computed Roche sphere radius that appear in the existing white dwarf system literature in Appendix~\ref{appendix}. 
We present our results in Section~\ref{s:results} and discuss them in Section~\ref{s:discussion} before summarizing our work in Section~\ref{s:summary}.

\section{Methods}\label{s:methods}

\begin{figure}
    \centering
    \includegraphics[width=\linewidth]{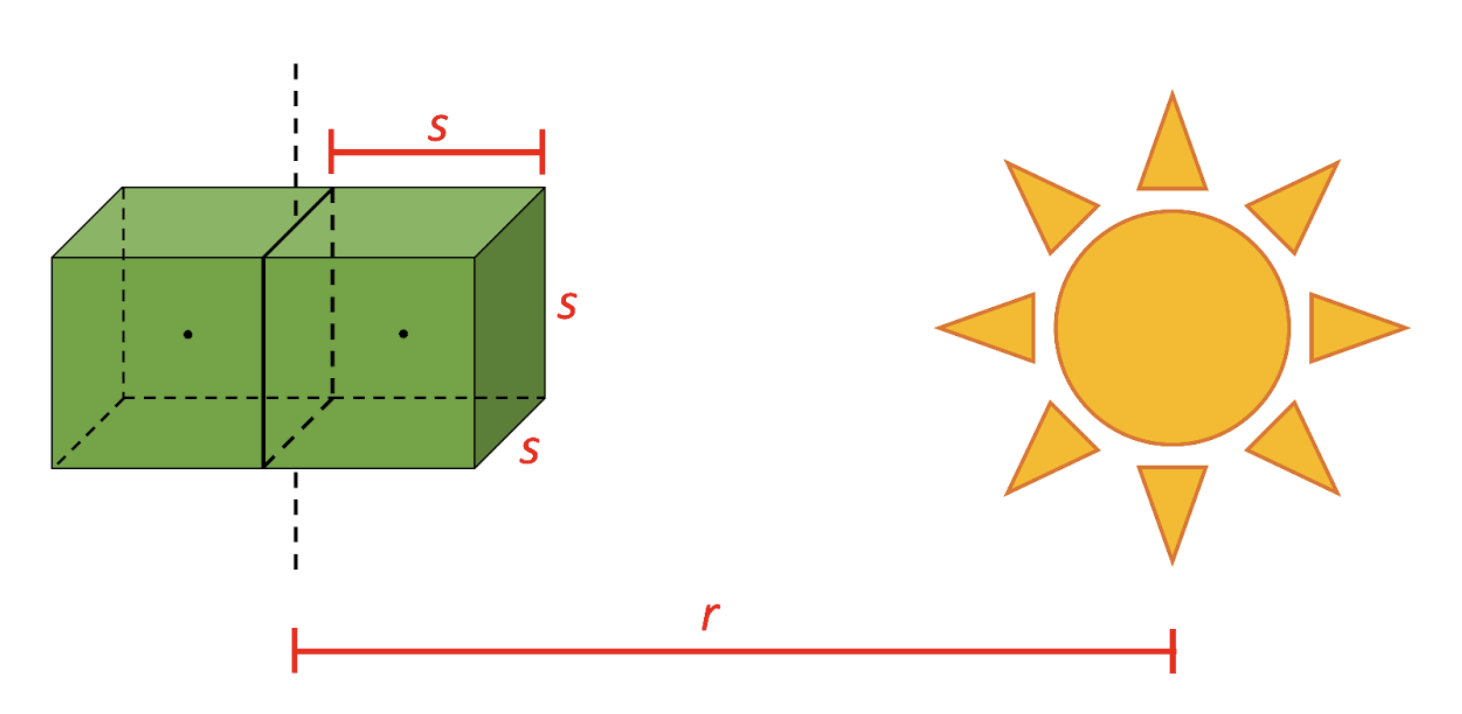}
    \caption{\textit{Model of tidally elongated asteroid}. 
    We model an inwardly scattered planetesimal as a rectangular prism comprising two cubes, which simplifies the problem whilst providing a conservative estimate of the minimum characteristic stable size.}
    \label{f:diagram}
\end{figure}

To understand how the characteristic size of orbiting tidal disruption fragments depends on material properties and distance to the star, we construct a simple model that compares the strength of the tidal force with the size dependence of material strength. 
We assume that tidal forces break up inwardly scattered planetesimals into minimally stable sizes, such that the tidal force is equal to the object’s strength. 
To easily compute these values, we build a simple model of a tidally elongated planetesimal as two cubes with side-length \jks{$s$} attached together (see Figure~\ref{f:diagram}).  
Although a rectangular equivalent to two cubes is an unlikely shape, it is nevertheless a reasonable approximation to an extremely elongated asteroid with an axial ratio of 0.5. 
Because Main Belt Asteroids have average axial ratios of 0.80 ± 0.04 \citep{Mommert:2018}, this model will make an asteroid that is more susceptible to tidal forces and therefore result in conservative estimates that err to smaller sizes, and therefore serve to strengthen the point of this work.

Geologic materials such as planets and planetesimals are not typically perfect monolithic crystals, but rather imperfect mixtures of minerals, rocks, and volatiles containing numerous flaws. 
Assuming these flaws are randomly distributed, tidal disruption events will cause planetesimals to split along a single plane of weakness \citep[e.g.,][]{Steckloff:2016}, and will continue to crumble until the debris reaches a minimally stable, characteristic size. 
To compute this size, we compare the tidal forces with cohesive strength and self-gravity. 
The magnitude of the cohesive force resisting tidal disruption is
\begin{equation}\label{eq:Fstrength}
F_{strength} = \sigma s^2 
\end{equation}
where $\sigma$ is the material cohesive (tensile) strength and $s$ is the length of each ``side'' of the cubes comprising this modeled planetesimal. 
The force of the planetesimals’ own gravity ($F_{gravity}$), which also opposes tidal disruption, is
\begin{equation}\label{eq:Fsg}
F_{gravity} = \frac{G m^2}{s^2} = \frac{G (\rho s^3)^2}{s^2} = G \rho^2 s^4
\end{equation}
where $G$ is the gravitational constant, $m$ is the mass of each ``cube'' (half the mass of the planetesimal), 
and $\rho$ is the density of the planetesimal. 

Meanwhile, tidal forces near periapse oppose the cohesive and self-gravitational binding forces. 
Assuming the asteroid spends sufficient time near periapse with its long axis pointed toward the white dwarf (maximizing the tidal forces), the magnitude of this peak tidal force is merely the difference in the gravitational attraction between each ``cube'' half of the planetesimal and the white dwarf due to their slightly different astrocentric distances
\begin{align}
F_{tide} &= F_{WD \,-\, near\,cube} - F_{WD \,- \,far\, cube}\\
&= \dfrac{GMm}{(r - s/2)^2} - \dfrac{GMm}{(r + s/2)^2} 
\end{align}
which simplifies to
\begin{equation}\label{eq:tide}
F_{tide} = \frac{GM\rho s^4 r}{(r^2 - s^2/4)^2}
\end{equation}
where $M$ is the mass of the white dwarf and $r$ is the astrocentric distance from the center of the white dwarf to the center of the asteroid.
We note that this derivation ignores the object's rotation, which would add an additional force that would aid in disrupting the object. 
Although the shortest known rotation period of a dwarf planet sized object is Haumea’s 3.9-hour period \citep{Rabinowitz:2006,Santos-Sanz:2017}, which is close to its disruption limit, the majority of such objects spin much more slowly; 
this renders the rotational component of the disruptive forces to be less important. 
Nevertheless, this biases our results to result in larger fragment sizes than would result if these rotational forces were included. 
Ultimately, this omission would change our results by no more than a small factor on the order of unity.

Combining these equations as a function of distance from the star, we can calculate the critical, minimally stable fragment size as a function of distance from the star:
\begin{align}
F_{tide} &= F_{strength} + F_{gravity}\\
\frac{GM\rho s^4 r}{(r^2 - s^2/4)^2} &=  \sigma s^2  + G \rho^2 s^4.
\end{align}
We solve this equation numerically to obtain the relationship between astrocentric distance ($r$) and the size of the minimally stable object size ($s$); 
we merely need to specify a white dwarf mass, and the planetesimal’s density and strength. 
See Appendix~\ref{appendix} for a comparison of this model with previously published models of tidal disruption around white dwarf stars.

\section{Results}\label{s:results}

\begin{figure*}
\begin{tabular}{cc}
\textbf{$\sigma$ = 10 Pa} & \hspace{-0.465in} \textbf{$\sigma$ =1 kPa} \\
\hspace{-0.65in}
\includegraphics[width=0.56\linewidth]{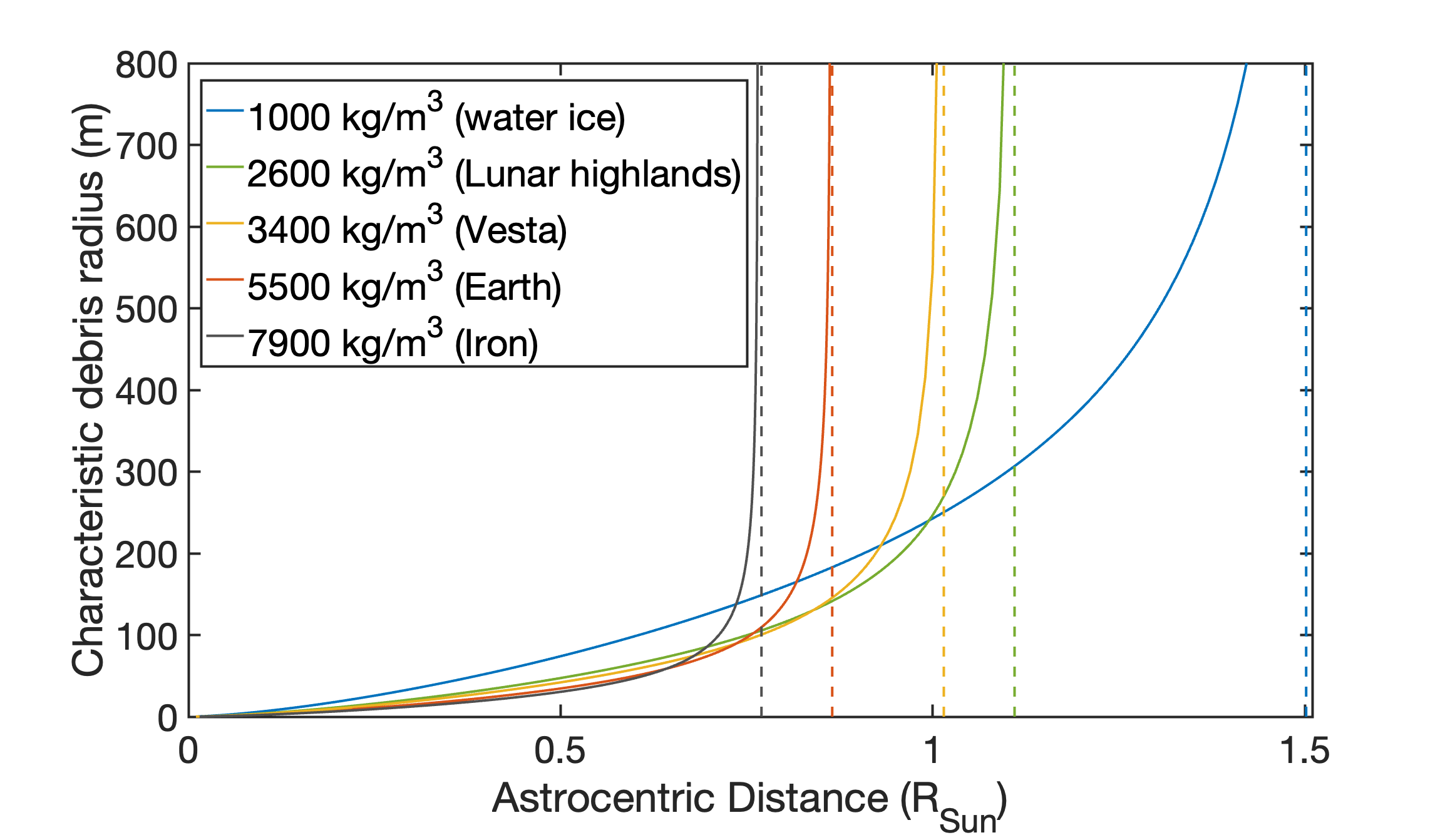} &
\hspace{-0.475in}
\includegraphics[width=0.56\linewidth]{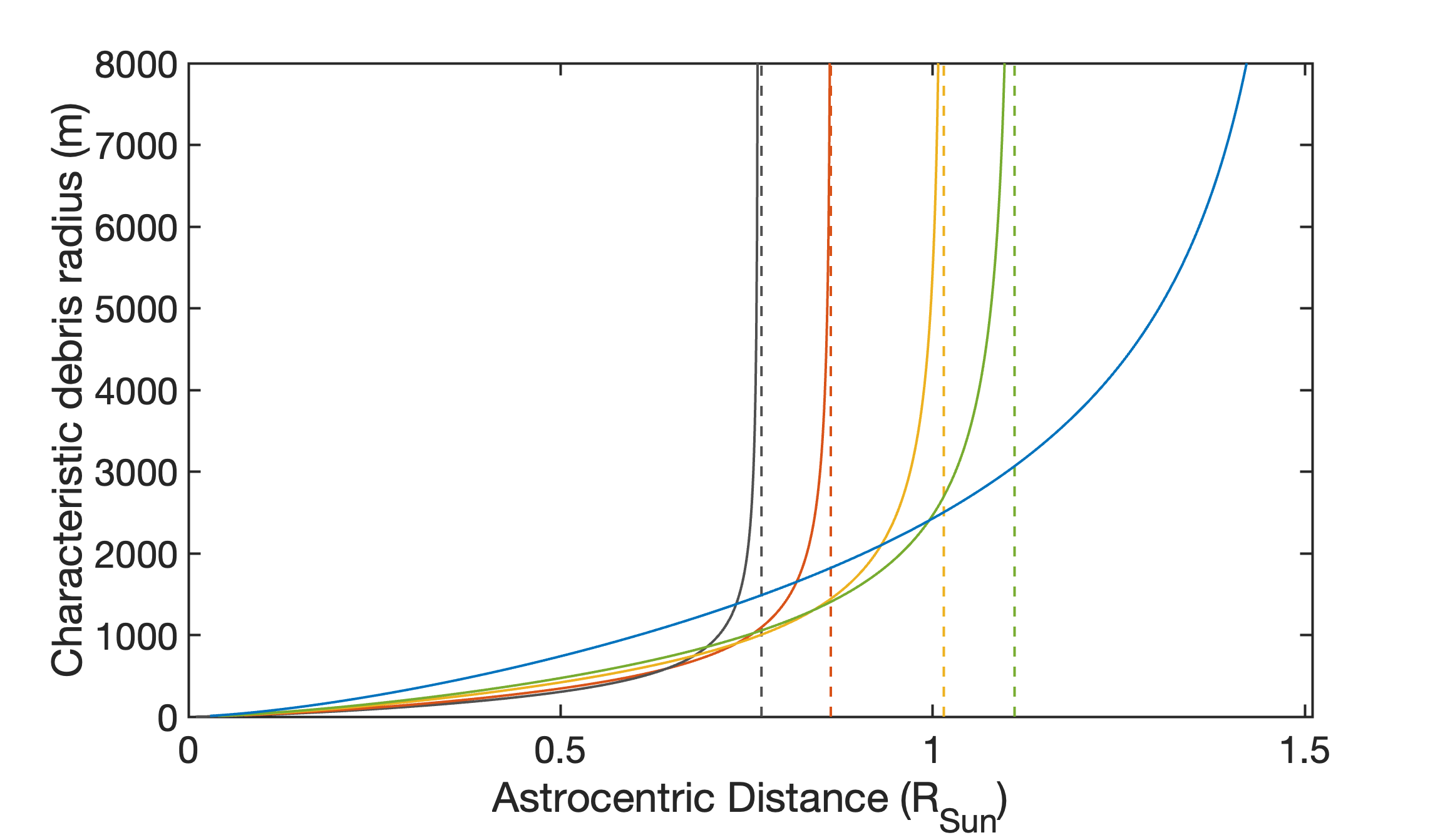} \\
\hspace{-0.55in}
\includegraphics[width=0.56\linewidth]{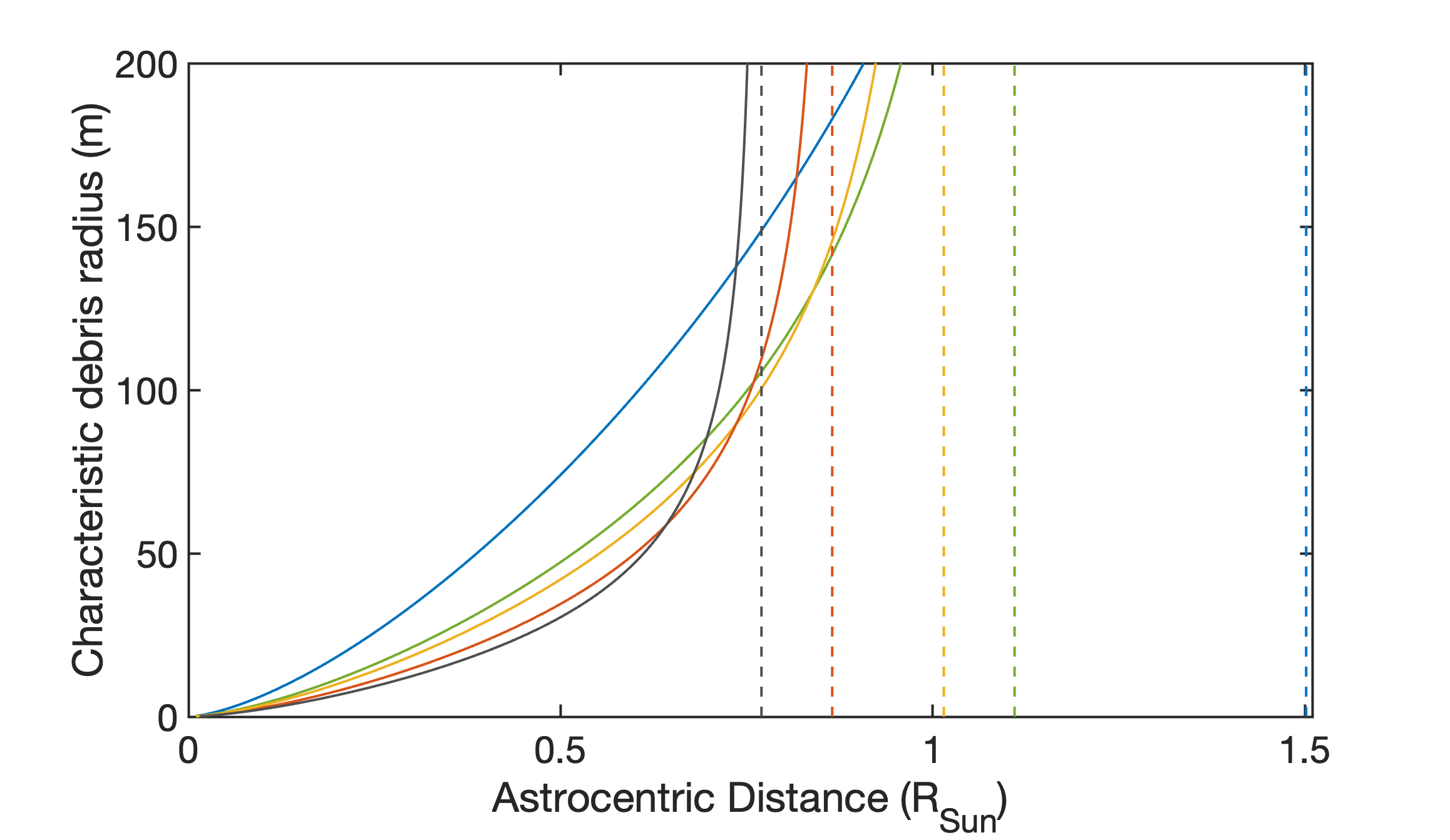} &
\hspace{-0.475in}
\includegraphics[width=0.56\linewidth]{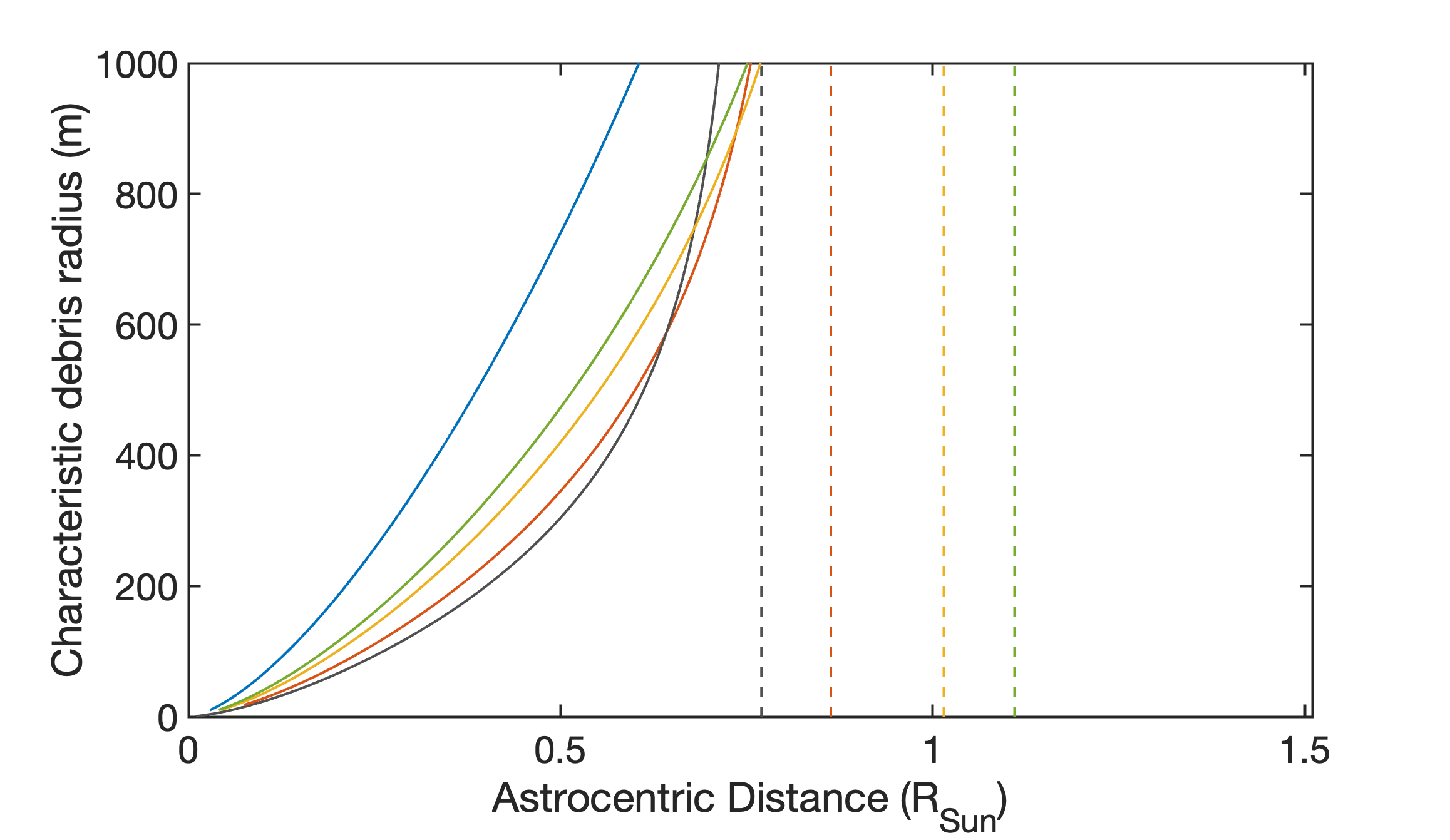} \\
\end{tabular}
    \caption{\textit{Characteristic debris size as a function of astrocentric distance and minimum cohesive strength.}
    \jks{We assume a 0.6 $M_{\sun}$ mass white dwarf}.
    The bottom panels are zoomed-in versions of the top panels. For minimum cohesive strengths of (left panels) 10 Pa and (right panels) 1 kPa, we find characteristic sizes on the order of $\sim$100 m and $\sim$1000 m respectively. Although the density of a material can strongly influence disruption behavior, the densities of real planetary materials typically tidally disrupt into debris with these characteristic sizes. 
    \label{f:results}}
\end{figure*}

Our main results are visualized in Fig. \ref{f:results}. 
In the figure, we compute the size of the debris as a function of astrocentric distance for strengths of 10 Pa and 1 kPa (note that greater strengths will result in larger stable debris sizes), and for densities characteristic of planetary objects of interest: 1000, 2600, 3400, 5500, and 7900 kg/m$^3$, representing the densities of ice, Vesta, Ceres, the Earth, and iron. 
Outside of the Roche limit, all sizes are stable because binding forces always dominate tidal disruption forces. 
However, inside of the Roche limit, maximum stable sizes drop rapidly to strength-dependent sizes on the order of $\sim$100 m for 10 Pa tensile strength and $\sim$1 km for 1 kPa tensile strength (see Figure~\ref{f:results}). 
Closer to the star, the stable size drops until it reaches zero. 
However, planetesimals are much more likely to be scattered onto orbits with larger perihelia closer to the Roche limit than they are to be scattered into orbits that pass even closer to the host star.  
Furthermore, extrapolating from \cite{Steckloff:2021}, the minimum perihelion distance at which any dust/debris is stable against sublimative destruction for the coldest white dwarf known to host a debris disk \citep{Debes:2019} is on the order of $\sim$0.1 solar radii ($R_{Sun}$). Thus, we focus on results in the region of disk stability, between the sublimation and Roche radii. 

\section{Discussion}\label{s:discussion}

Tides cannot disrupt exoplanetesimals into fragments larger than $\sim$0.1 - 1 km, regardless of the size of the parent body. 
Consequently,
\begin{itemize}
    \item These values place a useful upper bound on the largely unknown size distribution in white dwarf debris discs, helping establish initial conditions for models.
    \item The outcome of the tidal disruption mechanism may be more easily distinguished from other disruption mechanisms\footnote{One method is rotational disruption \citep{Veras:2020,Veras:2022,Veras:2026}, another is catastrophic disruption \citep{Veras:2025} and a third is thermal cracking \citep{Shestakova:2023}.} because those other mechanisms are not necessarily subject to the same physics illustrated in Figure~\ref{f:diagram}.
    \item These values affect the resulting collisional and radiative evolution of the system.
\end{itemize}
On this last point, most of the mass of the tidally disrupted debris will be in the $\sim$0.1 - 1 km fragments. 
These fragments are too large to be affected significantly by Poynting-Robertson drag. 
Hence, between the stages of tidal disruption and inward radiative drag into the sublimative zone, there should be a collisional phase, helping to affirm suggestions from recent observations \citep{Swan:2021,Noor:2025,Rogers:2025}.

How other radiative effects, such as the Yarkovsky and YORP effect, alter this picture is not yet clear. 
Figure 1 of \cite{Veras:2022} illustrates that both effects may play a role in this size range. 
However, these effects may act over longer timescales than the observations suggest; \cite{Aungwerojwit:2024} showed that the high level of dynamical activity in the WD 1145+017 system \citep{Vanderburg:2015} ceased completely within one decade, and activity levels in ZTF J1944+4557 ceased within one year \citep{Guidry:2025}.
Moreover, if the thermal environment is sufficiently warm to sublime volatiles, then sublimative torques can spin these objects up as well \citep{Steckloff:2016,Veras:2026}. 
Upon tidal disruption, the resulting fragments can spin up to disruption, until they have fragmented down into dust-sized debris susceptible to radiative forces \citep[e.g.][]{Steckloff:2016}, such as Poynting-Robertson drag.  

However, radiative and sublimative processes require a white dwarf to be sufficiently luminous (i.e., young) to drive rotational disruption, which weakens as the star ages. 
Nevertheless, dusty debris disks are present around very old white dwarfs, such as WD LSPM J0207+3331, a cool $\sim$5900K, $\sim$3 Gyr old white dwarf \citep{Debes:2019, LeBourdais:2025} or WD J2317+1830 a 4557 K, 6.4 Gyr old \citep{Bergeron:2022}; similarly, planetesimal pollution is observed within WD J2147-4035 (temperature of 3050 K, 10 Gyr cooling age) and WD J1922+0233 (3340 K, 9 Gyr cooling age) implying recent/ongoing accretion of planetary material \citep{Elms:2022}.
This demonstrates that non-radiative processes, such as collisional grinding, remain active long after purely thermally-driven processes shut down. 
Regardless, tidal breakup is merely an initial step in generating dust for debris disks.

The breakup of planetesimals into $\sim$0.1-1~km fragments is consistent with solar system observations. 
In 1993, Comet Shoemaker-Levy 9 famously broke into fragments after traveling within Jupiter’s Roche radius, and these fragments subsequently collided with Jupiter itself one after another. 
This provided two methods to estimate the size of the fragments. 
First, analysis of HST observations pre-impact showed $\sim$1 km fragments \citep{Weaver:1994}. 
Later, analysis of the impact debris fields themselves suggest $\sim$0.1 km fragments \citep{Knacke:1994}.

\begin{figure*}
\begin{tabular}{cc}
\textbf{$\sigma$ = 10 Pa} & \hspace{-0.465in} \textbf{$\sigma$ =1 kPa} \\
\hspace{-0.65in}
\includegraphics[width=0.56\linewidth]{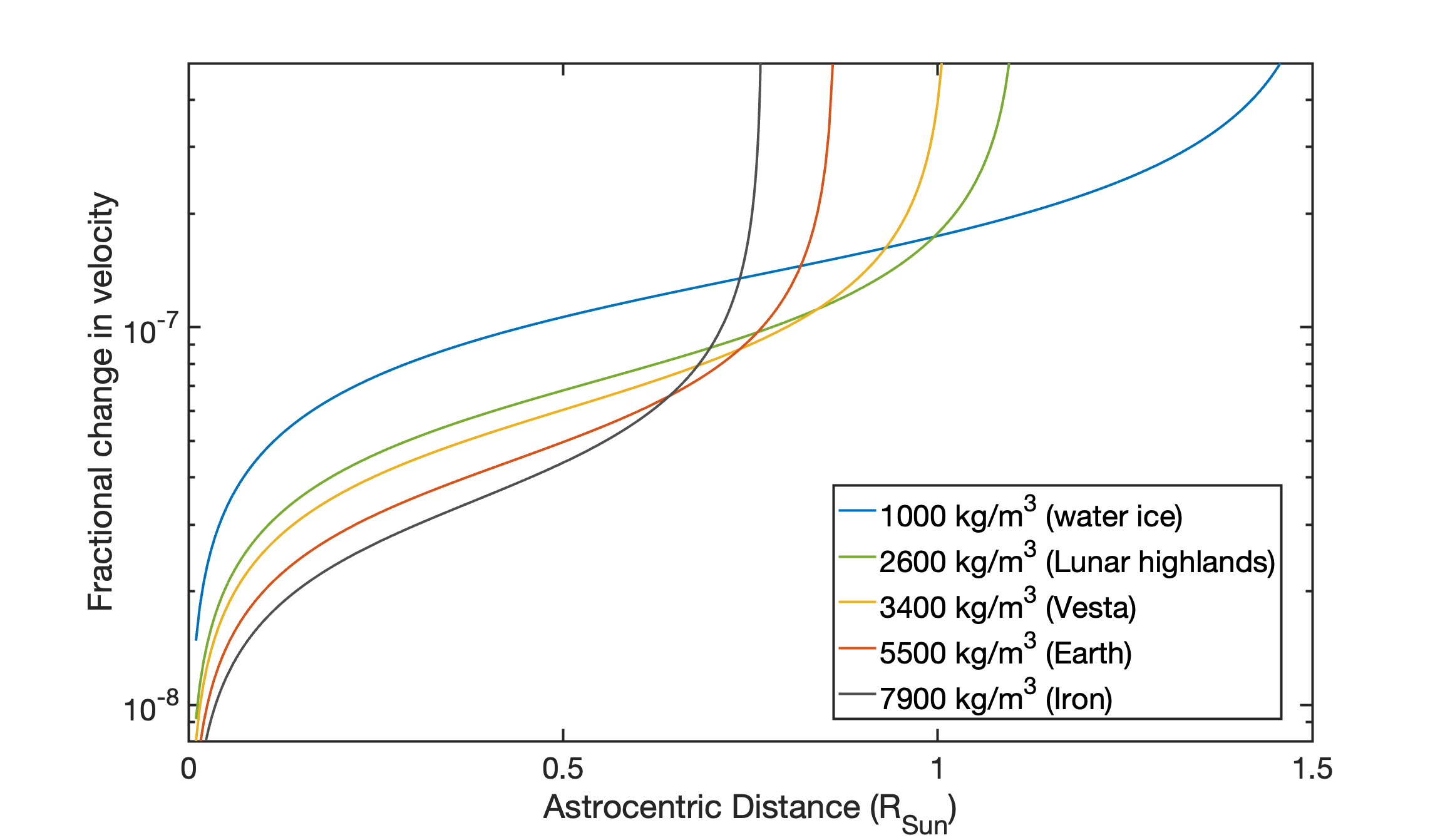} &
\hspace{-0.475in}
\includegraphics[width=0.56\linewidth]{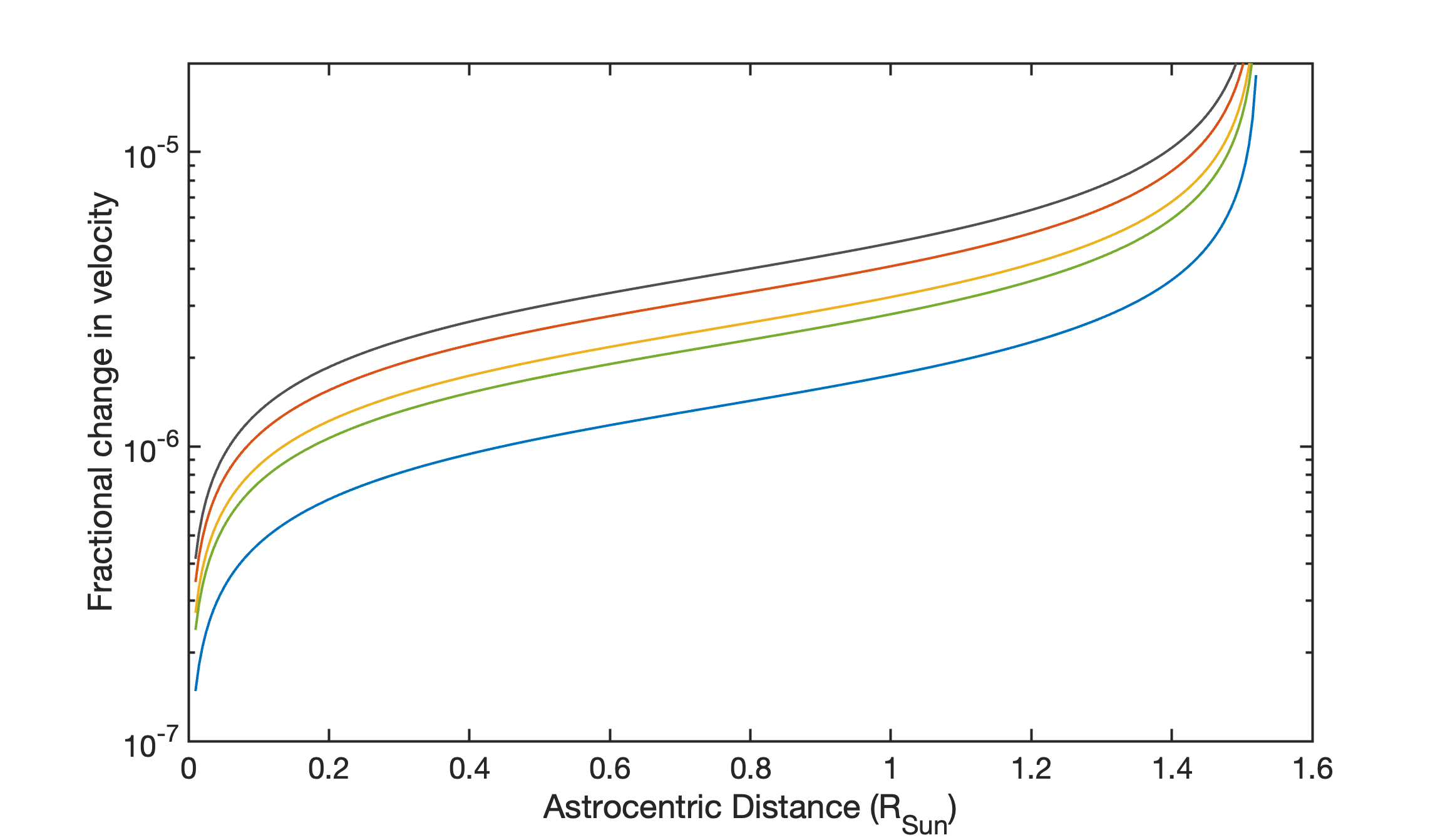} \\
\end{tabular}
    \caption{\textit{Change in fragment orbital velocity upon splitting as a function of astrocentric distance and minimum cohesive strength.}
    \jks{We assume a 0.6 $M_{\sun}$ mass white dwarf. We use the vis-viva equation to compute the fractional changes in the velocity of the fragments, which tend to be very small (on the order of 1 part per million). These tiny changes suggest that the fragments will remain in nearly identical orbits.}}
    \label{f:deltav}
\end{figure*}

Closer to the Sun, a large icy planetesimal passed within the Sun’s Roche radius and fragmented, to produce a trail of icy debris; this debris is now observed as the Kreutz Sungrazers group of comets \citep[see, e.g., review by][]{Jones:2018}. 
Although the majority of the known Kreutz sungrazers are tiny fragments, the majority of the mass of the comets resides in the large $\sim$1-10 km members of the group \citep{Knight:2010}, such as comet C/2026 A1 (MAPS) \citep{Zhang:2026}, or Comet Ikeya-Seki \citep{Knight:2010}. 
Thus, the tidal breakup of an eccentric, icy exoplanetesimal would likely create a structure similar to that of the Kreutz Sungrazers. 
As such, white dwarf stars hosting debris disks may be ideal candidates to search for signatures of exocomets.

This tidal disruption process suggests that white dwarfs with debris disks may host hidden dynamical structures, such as eccentric fields of fragments/debris. 
Although these primary fragments are likely too small to be directly detected, collisional grinding or rotational bursting will produce dust that will (at least initially) remain within the same orbit as the fragments, making these debris fields longer lived and more persistent; although out of phase, each disrupting fragment should produce a debris cloud with the same orbital period. Nevertheless, detection of a debris field on a highly eccentric orbit requires long-term, persistent observation to opportunistically detect such a recurrent, long-period debris cloud \citep{Vanderbosch:2020}. Nevertheless, short-period eccentric disks of gas debris have been detected, particularly with Doppler Tomograms \citep{Manser:2016,Manser:2021}. 

\jks{Imposing a minimum strength for tidally disrupted planetesimals has other consequences. One consequence is how the velocity dispersion of the fragments differs from the purely strengthless case \citep{Veras:2014,Veras:2017,Malamud:2020,Brouwers:2022}.}

\jks{To investigate, we computed the fractional change in the orbital velocities of the fragments upon breakup for non-rotating asteroids, and report the results in Figure \ref{f:deltav}.  In all cases, we find that the fractional change in the velocity of the fragments upon separation is on the order of one part per million to one part per 100 million. Such tiny changes in speed result in negligible changes to the orbits of the fragments, which remain in practically the same orbit as prior to disruption.}

\jks{Nevertheless, we see interesting structure in these plots due to the effects of strength, differentiating these results from previous studies. Internal strength enables objects to penetrate into the rubble-pile Roche limit before fragmenting.  As such, one might expect the velocity dispersion to increase as the object approaches the white dwarf.  However, the opposite is true; variations in orbital speed of the fragments decrease with \textit{decreasing} distance from the white dwarf.  The reason is that the sizes of the fragments decrease as the breakup location approaches the white dwarf, because the increasing tidal forces break the asteroids into ever smaller fragments. As these fragments get smaller, the relative spacing between the two daughter fragments decreases, causing their Keplerian orbits to become ever more similar. In contrast, further away from the white dwarf, the larger possible fragment sizes enable greater orbital separation between the daughter fragments, enabling a larger difference in their Keplerian orbital velocities.}

\jks{Thus, accounting for strength fundamentally changes the orbital behavior of the population of fragments, 
inhibiting their orbital spread. A narrower radial confinement of debris compared to the strengthless case may reduce the collisional timescale of the fragments as they subsequently spread due to Poynting-Robertson drag and general relativistic precession; future modelling may better quantify this idea.}
\smallskip
\section{Summary and Conclusions}\label{s:summary}

Understanding the progression from an intact minor planet to a sea of dust that is accreted onto a white dwarf is an outstanding challenge. In this Letter, we demonstrate a sharp constraint on the first step of this process.

We have considered the effects of strength, self-gravity, and tidal forces on white-dwarf minor planets to determine the characteristic size of the fragments resulting from crossing the Roche limit. 
All planetesimals -- even rubble pile asteroids held together by only Van der Waals forces -- possess sufficient cohesive strength to significantly affect the size frequency distribution of the fragments upon tidal disruption.

We consider planetesimal densities from 1000--7900 kg/m$^3$ (i.e., from ice to iron) and find that the majority of the parent body mass remains within $\sim$0.1--1 km fragments; this characteristic size range is consistent with tidally disrupted planetesimals in the solar system (i.e., Comet Shoemaker-Levy 9 or the Kreutz Sungrazers). 
Consequently, models should consider the evolution of these fragments as a separate precursor step to the eventual creation of dust. 
This step may feature collisional grinding and/or rotational disruption. 
Our prediction that fragments of these sizes are commonly formed in white dwarf planetary systems may affect disk formation and lifetimes, while helping to constrain the largely unknown size distribution of the debris closely orbiting white dwarfs.



\begin{acknowledgments}
\jks{We thank the reviewer for helpful comments which have improved the manuscript.} JKS would like to thank John Debes, Melinda Soares-Furtado, Andrew Vanderburg, David Nesvorny, and Jason Steffen for valuable conversations that significantly improved and constrained this work and its scope. 
JKS and KV were supported by NASA award 80NSSC20K0267. 
\end{acknowledgments}


\begin{appendices}
\section{Comparing with Published Models}\label{appendix}

Our numerical model solves for characteristic fragment size from a force balance. Many other published models also employ force balances, but use them instead to solve for the white dwarf Roche radius.
In this appendix, we compare our numerical with a selection of models, to demonstrate their similarity under relevant conditions/circumstances.

\cite{Li:2025} derive a force balance at the Roche limit (their Equation 8) of 
\begin{equation}
\frac{2GM}{r^3} + \omega^2 R = \frac{Gm}{R^2} + \frac{\sigma \pi R^2}{m},		
\end{equation}
where $G$ is the gravitational constant,  $M$ is the mass of the central star, $r$ is the astrocentric distance, $\omega$ is the angular velocity of the rotating planetesimal, $R$ is the planetesimal radius, $m$ is the planetesimal mass, and $\sigma$ is the planetesimal cohesive strength. 
From this relationship, they derive the location of the Roche limit
\begin{equation}
    r_{Roche} = \left( \frac{2GM}{\frac{4}{3} \pi G \rho + \frac{3 \sigma}{4 \rho R^2} - \omega^2} \right)^{1/3}.
    \label{LiEq}
\end{equation}
If we assume  that the planetesimal is not rotating, then Eq. (\ref{LiEq}) simplifies to 
\begin{equation}\label{eq:li}
    r_{Roche} = \left( \frac{2GM}{\frac{4}{3} \pi G \rho + \frac{3 \sigma}{4 \rho R^2}} \right)^{1/3}.
\end{equation}
The expression in Eq. (\ref{eq:li}) is equivalent to our numerical model if we consider the forces at the Roche limit and combine Equations~\ref{eq:Fstrength}, \ref{eq:Fsg}, and~\ref{eq:tide} such that
\begin{align}
F_{tide} &=  F_{gravity} + F_{strength},\\
\frac{GM\rho s^4 r}{(r^2 - s^2/4)^2} &=  G \rho^2 s^4  + \sigma s^2,  \\
\frac{GM s^2 r}{(r^2 - s^2/4)^2} &=  G \rho s^2  + \frac{\sigma}{\rho}.
\end{align}
Because $r \gg s$, the $r^4$ terms dominate the denominator on the left-hand side of the equation. 
Thus, we can simplify this expression to:
\begin{equation}
\frac{GM s^2}{r^3} =  G \rho s^2  + \frac{\sigma}{\rho}.
\end{equation}
We can then solve for $r$, which is really $r_{Roche}$, yielding:
\begin{equation}\label{eq:roche}
    r_{Roche} = \left(\frac{GM}{G \rho + \frac{\sigma}{\rho s^2}} \right)^{1/3}.
\end{equation}
We can then directly compare Equation~(\ref{eq:roche}) with the non-rotating case from \cite{Li:2025} (Equation~\ref{eq:li}); we find that the two equations have the same functional form, with minor differences accounted for by the geometric choices of planetesimal shape.

In a similar model, \cite{Shestakova:2023} also solve the force balance at the Roche limit -- but neglect self-gravity -- with the goal of estimating the critical size at which objects become unstable and break up. They derive the relationship:
\begin{equation}\label{eq:S26}
    R \le \frac{3}{4}\left(\frac{r}{R_{*}}\right)^{3/2} \sqrt{\frac{\sigma}{2 \pi G \rho_{*}\rho}}
\end{equation}
where $R_{*}$ and $\rho_{*}$ are the radius and density of the central star. 
If we similarly neglect self-gravity at the Roche limit, then our Equation~(\ref{eq:roche}) becomes
\begin{equation}\label{eq:gf-roche}
    r_{Roche} = \left(\frac{GM \rho s^2}{\sigma} \right)^{1/3}.
\end{equation}
We can solve for s, yielding
\begin{equation}
s = \sqrt{ \frac{r_{Roche}^3 \sigma}{G M \rho} }
\end{equation}
and  we note that mass of the star $M$ is that of a sphere of radius $R_{*}$ and density $\rho_{*}$ to {\DV obtain}
\begin{equation}
    s = \sqrt{ \frac{3 r_{Roche}^3 \sigma} {4\pi G R_{*}^3 \rho_{*} \rho} }.
    \label{Eq:ssEquiv}
\end{equation}
Equation (\ref{Eq:ssEquiv}) can be rearranged to
\begin{equation}
    s = \sqrt{\frac{3}{2}} \left( \frac{r}{R_{*}}\right)^{3/2} \sqrt{\frac {\sigma}{2\pi G \rho_{*}\rho} },
\end{equation}
which has the same functional form as found in Equation~(\ref{eq:S26}) \citep{Shestakova:2023}, albeit with a different leading coefficient, again due to geometry. This comparison reveals that for a given size ($s$), we need a larger distance from the star due to the smaller cross-sectional area (and thus, strength) required to hold the body together, consistent with expectations.

We also compare our model to the model of \cite{Zhang:2021}, who also assume negligible gravity at the Roche limit
\begin{equation}
    r_{Roche} = \left( \frac {\sqrt{3}}{4\pi}  \right)^{1/3}
    \left(  \frac{5k}{4\pi GR^2\rho^2} +s'    \right)^{-1/3}
    \left( \frac{M}{\rho} \right)^{1/3},
\end{equation}
where $k$ and $s'$ are constants related to the internal friction angle $\phi$  and cohesive strength $\sigma$ of the material comprising the planetesimal:
\begin{align}
    s' &= \frac{2\sin\phi}{\sqrt{3}(3 - \sin\phi)},\\
    k &= \frac{6\sigma\cos\phi}{\sqrt{3}(3 - \sin\phi)}.
\end{align}
Given that most materials have an angle of internal friction of $\sim$30$^\circ$ \citep{Melosh:2011}, these expressions for $s'$ and $k$ become
\begin{align}
    s' &= \frac{2}{5\sqrt{3}},\\
    k &= \frac{6\sigma}{5},
\end{align}
and thus
\begin{equation}
    r_{Roche} = \left( \frac {\sqrt{3}}{4\pi}  \right)^{1/3}
    \left(  \frac{6\sigma}{4\pi GR^2\rho^2} +\frac{2}{5\sqrt{3}}    \right)^{-1/3}
    \left( \frac{M}{\rho} \right)^{1/3}.
    \label{Eq:VdW}
\end{equation}
Even for Van der Waals attraction within small rubble-pile asteroids of $\sim$30 Pa \cite{Sanchez:2014}, the strength term dominates the constant term in the middle parenthesis for $\sim$100 m asteroids, and is at worst comparable to it for $\sim$1 km asteroids.  
Thus, we can simplify Equation~(\ref{Eq:VdW}) for realistic scenarios in the large size limit to 
\begin{equation}
    r_{Roche} = \left(\frac{\sqrt{3}GM\rho R^2 } {6\sigma} \right)^{1/3}.
    \label{Eq:LL}
\end{equation}
If we compare Equation~(\ref{Eq:LL}) to our result in the gravity-less regime (Equation~\ref{eq:gf-roche}), we find that our result, due to the geometry of the chosen system, requires our non-spherical object to remain further out at a distance of $\sim$150\% of the value of \cite{Zhang:2021} to remain stable, consistent with expectations.

We further compare our results to the Roche radius calculation of \cite{O'Connor:2020}, who also considered objects with strength (assuming the strength of rock or iron) and self-gravity with the expression
\begin{equation}
    r_{Roche} 
    = K'R_{*} \left(\frac{G\rho_{*}}{\frac{\sigma}{\rho s^2} + G\rho} \right)^{1/3},
    \label{Eq:OcoEq}
\end{equation}
where $K'$ is a factor that depends on the shape of the planetesimal (assumed to be $\sim$2).
Again, assuming a spherical white dwarf of constant density, we can re-express Equation~(\ref{Eq:OcoEq}) as:
\begin{equation}
    r_{Roche} = \left( \frac{6}{\pi} \right)^{1/3}
    \left(\frac{G M }{\frac{\sigma}{\rho s^2} + G \rho} \right)^{1/3}.
\end{equation}
We can directly compare this with our result (Equation~\ref{eq:roche}). We find that these results are systemically $\approx$24\% larger than our values, which is in good agreement given the relatively generic geometries considered.

Our penultimate comparison is to the aspherical tidal fragmentation condition described in \cite{Bear:2015} (their Equation 8):
\begin{equation}
    \frac{2 G M}{r_{roche}^3} = \frac{4 \pi}{3}G \rho + \frac{\sigma}{\rho R^2}.
    \label{Eq:BeSo}
\end{equation}
Equation (\ref{Eq:BeSo}) can be rearranged to:
\begin{equation}
    r_{roche} = \left( \frac{2 G M}{ \frac{4\pi}{3}G \rho  + \frac{\sigma}{R^2\rho}} \right)^{1/3},
\end{equation}
which is of a similar functional form as our Equation~\ref{eq:roche}, albeit for a spherical planet.  
However, the gravity term included is $\sim$4 times larger than in our equation, as they use the surface acceleration of the planet, rather than the mutual gravitational attraction between the two hemispheres that split along the same plane along which strength is assumed to apply.

Our final comparison is to \cite{Brown:2017}. 
Taking their Equations 22-25, at the Roche radius
\begin{equation}
    \frac{\sigma}{G \rho^2 s^2} + 1 = \frac{M}{2 \rho r^3}
\end{equation}
which yields
\begin{equation}
    r_{Roche} = \left(\frac{1}{2}\right)^{1/3} \left( \frac{G M}{\frac{\sigma}{\rho s^2} + G\rho}  \right)^{1/3}.
\end{equation}
Their chosen geometry does not differentiate between spheres and cubes, but yields a  Roche radius $\approx$79\% of ours (Equation~\ref{eq:roche}).

\end{appendices}

\end{document}